\documentclass[preprint2]{aastex63}

\usepackage{amsmath}
\usepackage{enumitem}
\usepackage{CJK}

\shorttitle{CMR in BSS Field}

\begin{document}
\begin{CJK*}{UTF8}{gbsn}

\title{Dense Gas Formation via Collision-induced Magnetic Reconnection in a Disk Galaxy with a BiSymmetric Spiral Magnetic Field}

\author{Shuo Kong (孔朔)}
\affiliation{Steward Observatory, University of Arizona, Tucson, AZ 85719, USA}

\begin{abstract}
Recently, a collision-induced magnetic reconnection (CMR) mechanism was proposed to explain a dense filament formation in the Orion A giant molecular cloud. A natural question is that whether CMR works elsewhere in the Galaxy. As an initial attempt to answer the question, this paper investigates the triggering of CMR and the production of dense gas in a flat-rotating disk with a modified BiSymmetric Spiral (BSS) magnetic field. Cloud-cloud collisions at field reversals in the disk are modeled with the Athena++ code. Under the condition that is representative of the warm neutral medium, the cloud-cloud collision successfully triggers CMR at different disk radii. However, dense gas formation is hindered by the dominating thermal pressure, unless a moderately stronger initial field $\ga5\mu$G is present. The strong-field model, having a larger Lundquist number $S_L$ and lower plasma $\beta$, activates the plasmoid instability in the collision midplane, which is otherwise suppressed by the disk rotation. We speculate that CMR can be common if more clouds collide along field reversals. However, to witness the CMR process in numerical simulations, we need to significantly resolve the collision midplane with a spatial dynamic range $\ga10^6$. If Milky Way spiral arms indeed coincide with field reversals in BSS, it is possible that CMR creates or maintains dense gas in the arms. High-resolution, high-sensitivity Zeeman/Faraday-Rotation observations are crucial for finding CMR candidates that have helical fields.
\end{abstract}


\section{Introduction}\label{sec:intro}
\end{CJK*}

Magnetic field is thought to be an important factor that slows down the formation of dense gas/clouds by resisting the gravitational collapse. However, in special cases, magnetic field may also favor the dense gas formation. For instance, recently, \citet[][hereafter K21]{2021ApJ...906...80K} found that a collision-induced magnetic reconnection (CMR) mechanism can create a dense filament that is wrapped by helical fields. Most importantly, their filament explained key features in their observational data. Their results implied a new class of dense gas in the interstellar medium (ISM).

\begin{figure*}[htb!]
\centering
\epsscale{0.55}
\plotone{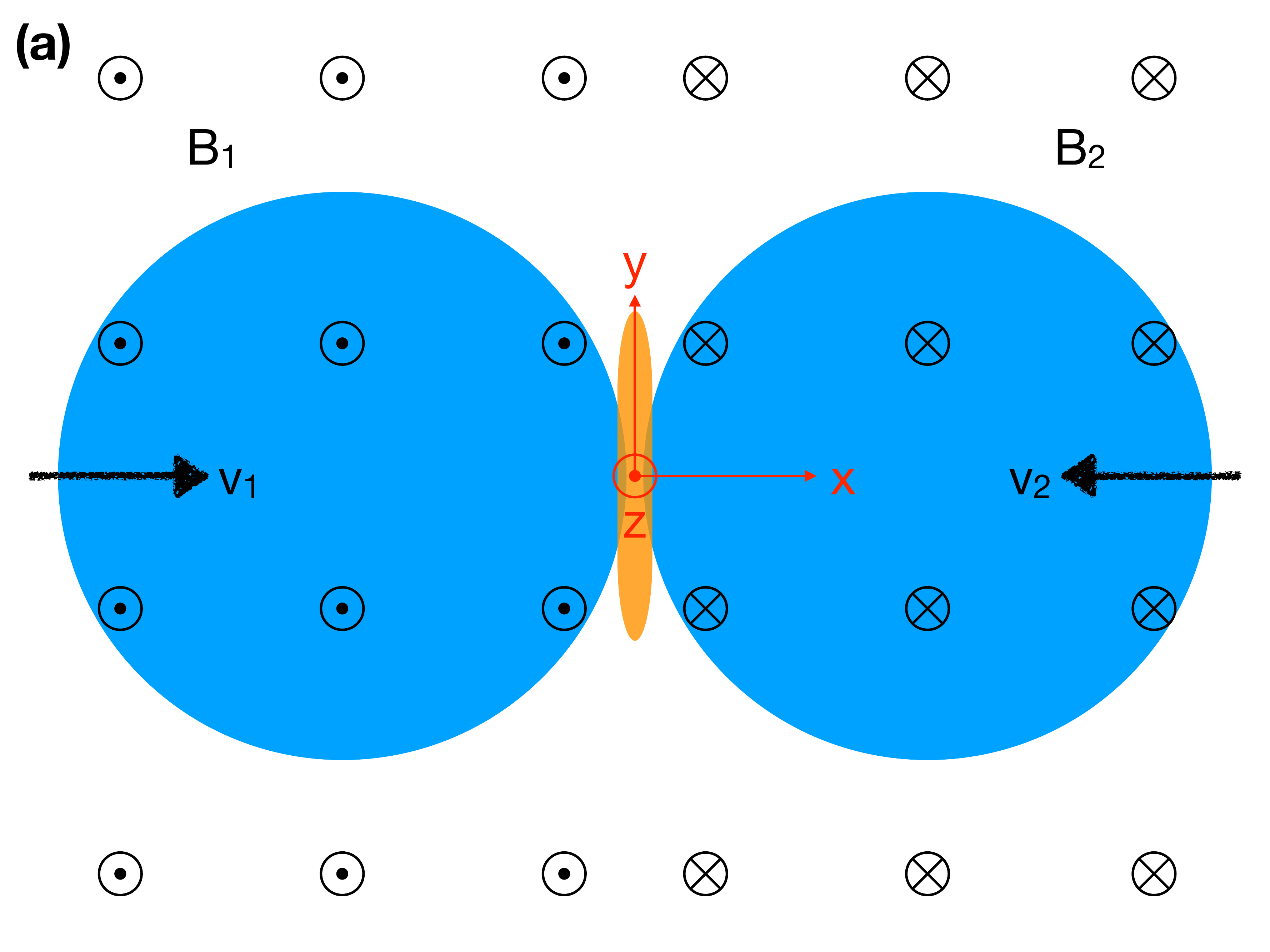}
\plotone{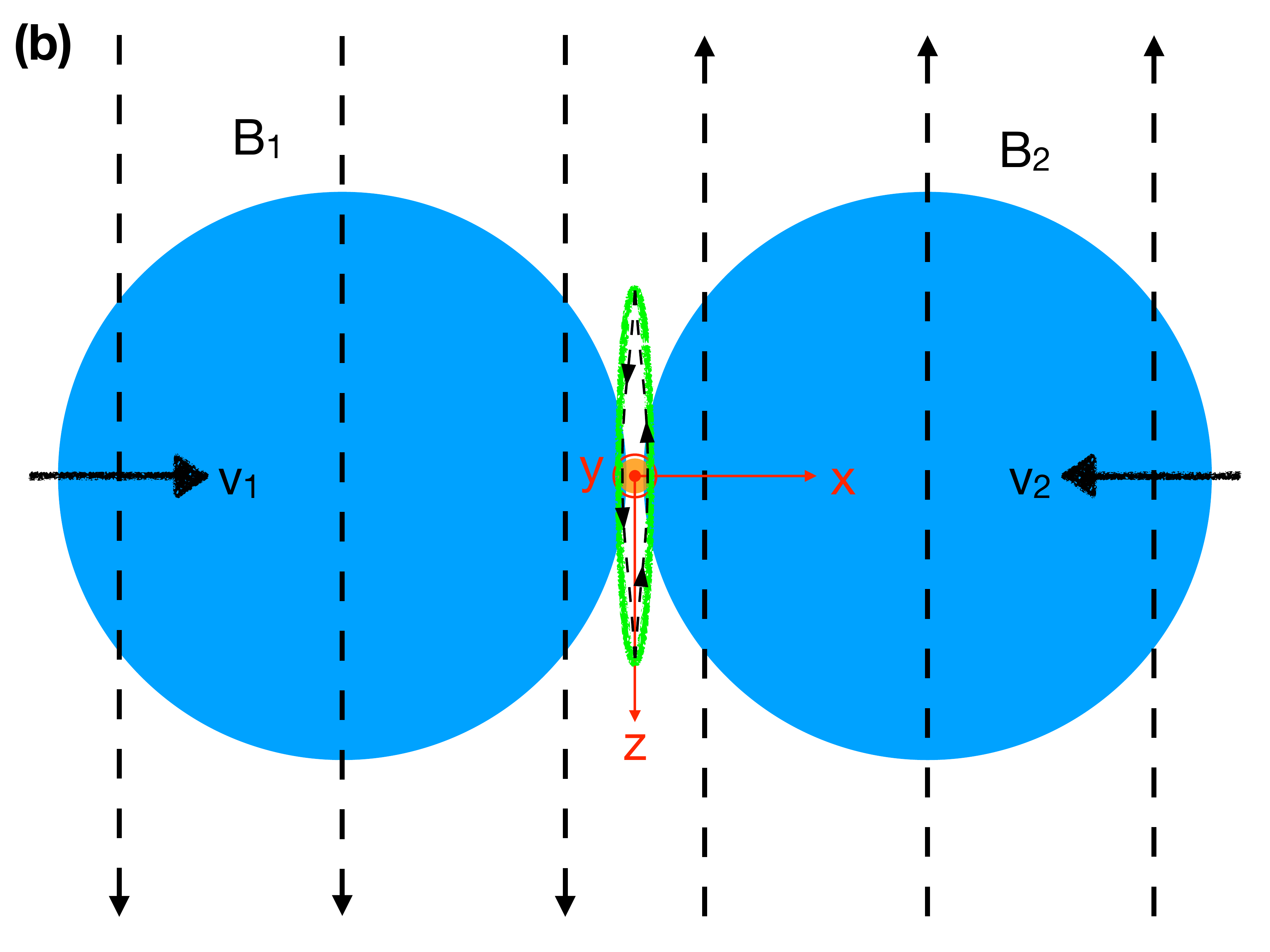}
\caption{
An illustration of CMR in two viewing angles. {\bf (a):} A view in the x-y plane at z=0. The Cartesian coordinate system (red) centers at the collision point. The x-axis points rightward and the y-axis points to the top. The z-axis points toward us as indicated by the red circle-point. The clouds have incoming velocities $v_1$ and $v_2$, respectively. The magnetic field points toward us (marked as black circle-points) for $x<0$ and away from us (marked as black circle-crosses) for $x>0$. After collision, the filament (orange) forms along the y-axis. {\bf (b):} A view in the z-x plane at y=0. In this view, the magnetic field is parallel to the plane of the sky. The y-axis points toward us as indicated by the red circle-point. After collision, the filament (orange) forms along the y-axis which points toward us. The green ellipse marks the location of the compression pancake if no magnetic fields. With antiparallel fields and CMR, the field reconnects at two ends of the pancake and forms a loop (black dashed arrow curve) enclosing the pancake. Due to the magnetic tension force, the pancake is squeezed into the central axis (y-axis) becoming a filament.
\label{fig:cmr}}
\end{figure*}

Figure \ref{fig:cmr} illustrates the CMR filament formation. In panel (a), we view the process from the side of the filament along the negative z-axis. Two clouds move along the x-axis and collide at the origin. On the left side of the x=0 plane, the magnetic field points toward us. On the other side, the field points away from us. After collision, the reversed field reconnects and forms field loops in the z-x plane which pulls the compression pancake into the central axis (the y-axis). As a result, a filamentary structure forms along the axis. In panel (b), we view the process in the z-x plane. In this projection, we are looking at the filament cross-section at the origin. The green ellipse represents the compression pancake and the black dashed arrow curve around the pancake denotes the reconnected field loop. The loop has a strong magnetic tension that pulls the dense gas in the pancake to the origin in each z-x plane. As a result, the filament (orange cross-section) forms along the y-axis. Essentially, the filament forms along the field symmetry axis that crosses the collision point. 

A natural follow-up question is that how common the CMR mechanism can be in our Milky Way. The reason for the question is that the prerequisite of antiparallel fields for CMR is seemingly rare, at least within giant molecular clouds \citep[however, turbulence may create copious field reversals, e.g.,][]{2016JPlPh..82f5301F,2018PhRvL.121p5101D}. On the other hand, cloud-cloud collision can be common in galaxies \citep[e.g.,][]{2017MNRAS.469..383J,2021PASJ...73S...1F,2021MNRAS.502.2238M}. But the magnetic field topology at the collision front is not well known. It would be useful to look for CMR conditions in an entire galaxy to assess the likelihood of CMR. 

For decades, there is evidence that our Milky Way has the so-called BiSymmetric Spiral (BSS) large-scale magnetic fields (\citealt{1983ApJ...265..722S}, hereafter SF83; \citealt{1994A&A...288..759H}, hereafter HQ94; \citealt{2017ARA&A..55..111H,2019Galax...8....4B}). With the increase of Galactocentric radius, magnetic fields reverse several times (see Figure \ref{fig:ic}). The field reversal naturally creates anti-parallel magnetic fields. If two clouds on both sides of the reversal run into each other, they will likely fulfill the initial conditions for CMR. Interestingly, HQ94 pointed out that field reversals coincide with spiral arms in the Galaxy. Is it possible that CMR contributes to the dense gas in spiral arms?

In this paper, we investigate the possibility of CMR in a disk galaxy with a BSS field that is established for our Milky Way (SF83,HQ94). The immediate goal is to check if CMR can be triggered within the disk. If so, how much dense gas is generated by CMR? In the following section, we describe the initial conditions for the fiducial model, including the gas disk and a modified-BSS field. In section \S\ref{sec:method}, we introduce the numerical method. In \S\ref{sec:results}, we present results from the numerical simulations. We then discuss the results and their implications in \S\ref{sec:discus}. Finally, \S\ref{sec:conclusion} will summarize and conclude the paper.

\section{Initial Conditions}\label{sec:ic}

\begin{figure*}[htb!]
\centering
\epsscale{1.1}
\plotone{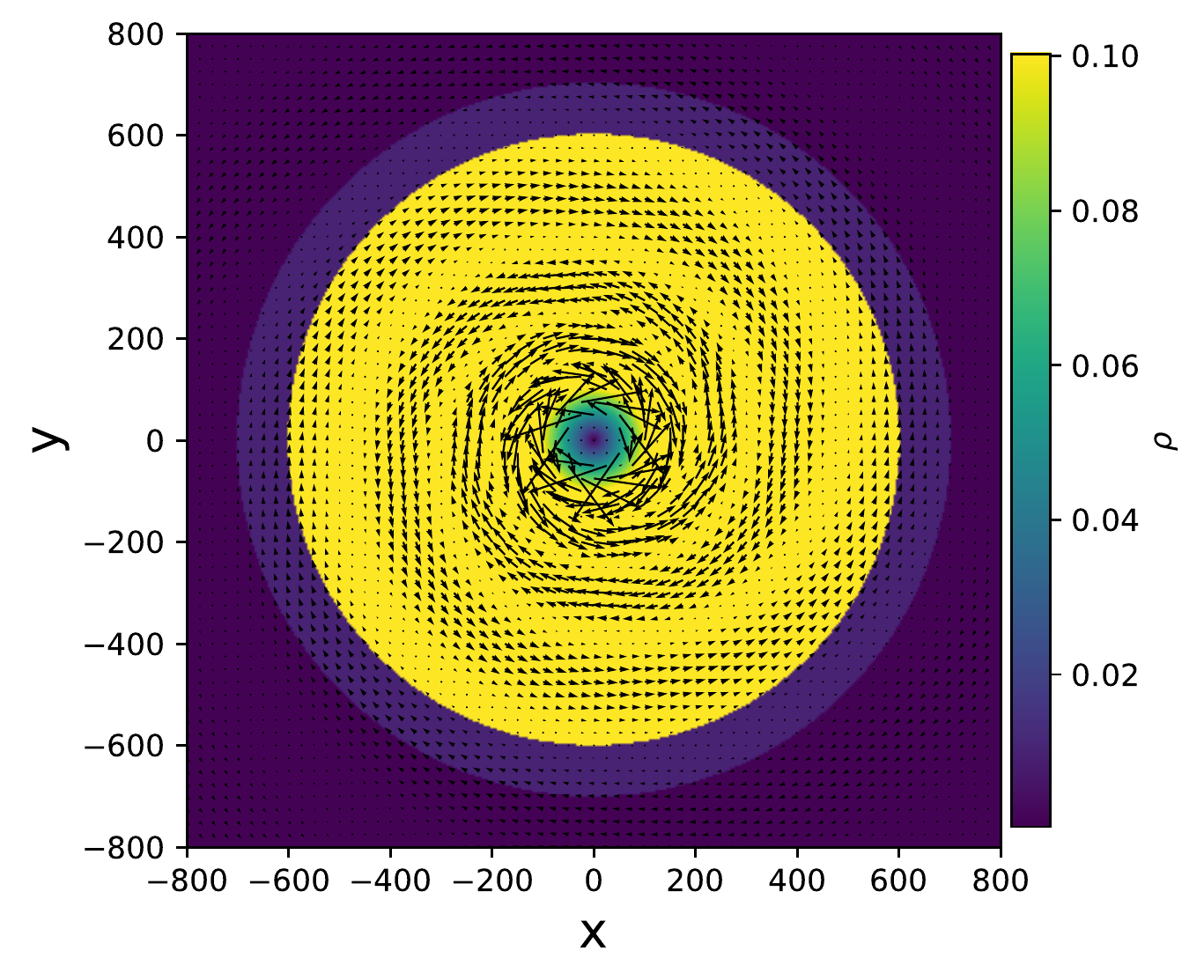}
\caption{
Initial conditions for the disk and the modified-BSS field (see \S\ref{sec:ic}). The color shows the density structure. The arrows show the magnetic field vectors. The arrow length is proportional to the field magnitude. The code unit is defined in \S\ref{sec:method}.
\label{fig:ic}}
\end{figure*}

For this exploratory study, we consider a simple two-dimensional, uniform gaseous disk\footnote{The uniform density is a simplified consideration. In reality, the turbulent ISM should develop a density distribution \citep[e.g.,][]{2010A&A...512A..81F} as well as a stochastic field \citep[e.g.,][]{2016JPlPh..82f5301F,2020MNRAS.499.4785K,2021MNRAS.502.2220S}. Future studies of CMR in realistic conditions are necessary, especially in 3D cases.}. As shown in Figure \ref{fig:ic}, the disk is in a two-dimensional domain of 16$\times$16 kpc. The main disk of interest is between radii $r$=1-6 kpc. This is a rather limited spatial region which saves computation cost because we want to resolve a scale of $\sim$0.05 pc (see below). To model the warm neutral medium (WNM), the uniform density is set to 3.84$\times$10$^{-24}$ g cm$^{-3}$ (corresponding to a number density of $n_H$ = 1.7 cm$^{-3}$). Within $r<1$ kpc, the gas density smoothly drops to a very low value at $r=0$. Between $r$ = 6 kpc and 7 kpc, an annulus of gas with a factor of 10 smaller density ($n_H$ = 0.17 cm$^{-3}$) co-rotates with the main disk. Beyond $r$ = 7 kpc, the space is filled with a static diffuse gas with $n_H$ = 0.017 cm$^{-3}$. For the WNM, we adopt an isothermal temperature of 8000 K, corresponding to a sound speed of $\sim$ 9 km s$^{-1}$.

To isolate the magnetohydrodynamics (MHD) effect and minimize effects due to any disk evolution, we put the disk in a static background gravitational potential that results in a steady-state flat rotation curve \citep{2008gady.book.....B}. For the two-dimensional disk, the potential is set as
\begin{equation}\label{equ:Phi}
\Phi = \frac{1}{2}v_{c,\infty}^2\ln\left(1+\frac{r^2}{r_c^2}\right),
\end{equation}
where $v_{c,\infty}$ is the rotation velocity when $r \gg r_c$. We adopt $v_{c,\infty}=200$ km s$^{-1}$ and $r_c=0.5$ kpc. The resulting rotation velocity $v_c$ is
\begin{equation}\label{equ:vel}
v_c = \frac{v_{c,\infty}r}{\sqrt{r_c^2+r^2}}.
\end{equation}
Because we focus on the gas evolution due to the interaction with magnetic fields. We do not include self-gravity in the gas disk. As shown in K21, the dense gas formation was solely caused by the reconnected field and self-gravity was negligible at least for the first 1 Myr. 

The initial magnetic fields follow the fitting results in SF83 and HQ94. Both studies used the Faraday Rotation Measures (RMs) of either Galactic or extragalactic radio sources to detect the line-of-sight magnetic field in our Milky Way. They found field reversals at spiral arms where the field is weak. The field is strong in the interarm region. A BSS field configuration was able to explain the observational features. HQ94 also attempted to fit a concentric ring model to the RMs. Their conclusion was that the BSS model gave a better fitting result. Both SF83 and HQ94 favored that the BSS field was primordial. Note that the field we discuss here is the large-scale field, neglecting local fields. 

\begin{figure}[htb!]
\centering
\epsscale{1.1}
\plotone{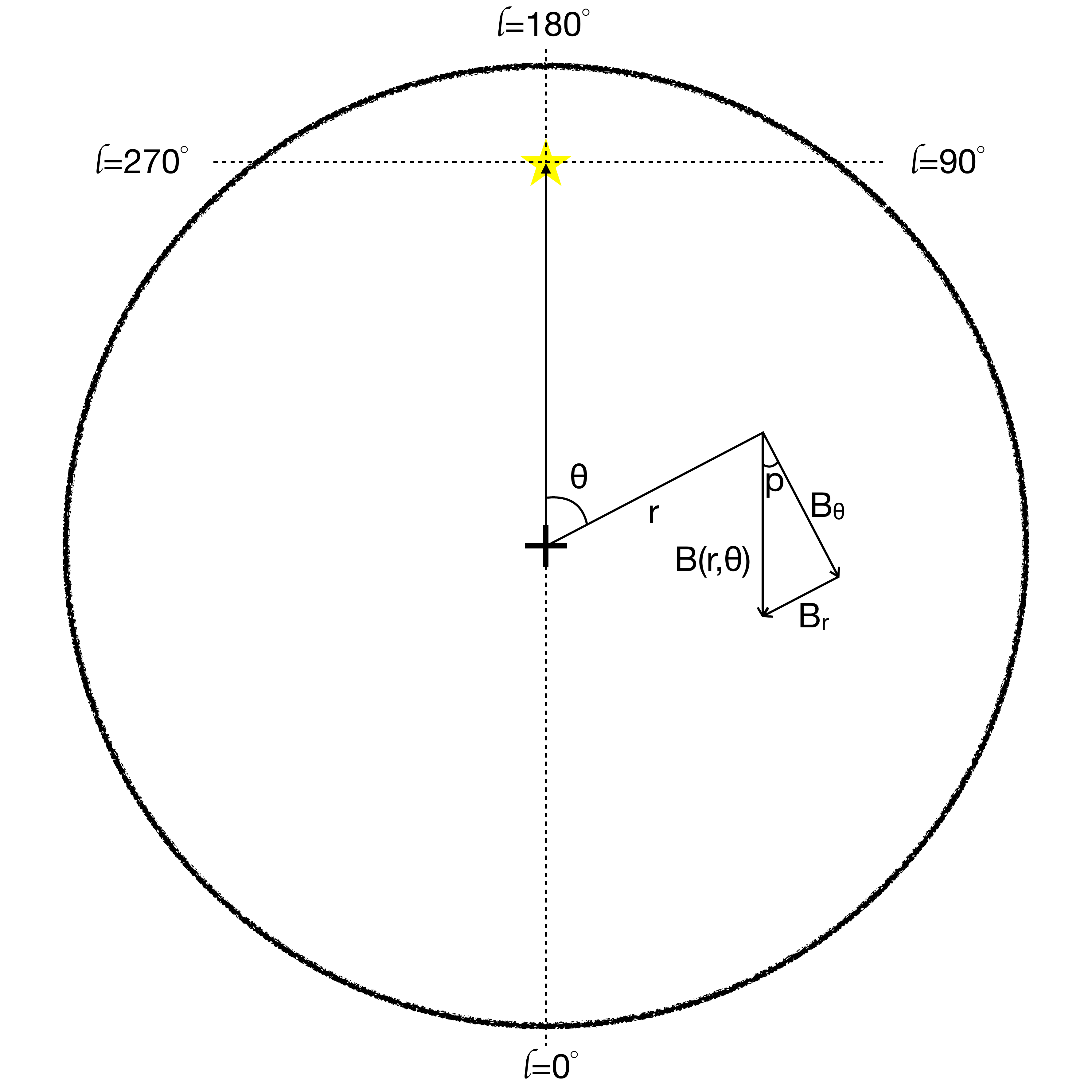}
\caption{
Coordinate system for the BSS field configuration. The big circle represents the Milky Way disk at 10 kpc. The yellow star roughly marks the location of the Sun. The central cross is the system center. The field strength at (r,$\theta$) is $B(r,\theta)$, which is split into the $r$ and $\theta$ components $B_r$ and $B_\theta$ by the pitch angle $p$. Following HQ94, the angle $p$ in the figure is negative. Note the actual size of the main disk in our models is 6 kpc to save computation cost (see \S\ref{sec:method}).
\label{fig:illus}}
\end{figure}

The specific mathematical expressions for the BSS field in HQ94, as will be shown in appendix \S\ref{app:bfield}, did not guarantee divergence-free in 2D. To reconcile their expressions without significant changes to the BSS model, we adopt the HQ94 expression but include the 1/$r$ factor in the leading coefficient (similar to the SF83 model A, but see \S\ref{app:bfield}). The final expression for the modified-BSS field is
\begin{equation}\label{equ:br}
B_r=B_0\frac{r_\odot}{r}\cos\left(\theta-\beta_p\ln\frac{r}{r_0}\right)\sin p,
\end{equation}
\begin{equation}\label{equ:bt}
B_\theta=B_0\frac{r_\odot}{r}\cos\left(\theta-\beta_p\ln\frac{r}{r_0}\right)\cos p,
\end{equation}
where $r$, $\theta$, $p$ are illustrated in Figure \ref{fig:illus}. $p$ is the pitch angle and is defined to be negative if it rotates clockwise from the tangential component $B_\theta$. The term $\beta_p$ is 1/tan$p$. The term $r_0$ is the location in the direction of $\theta=0$ that the field first reaches the local maximum outside the solar circle (HQ94). $r_\odot$ is the Galactocentric radius of the Sun. As will be shown in \S\ref{app:bfield}, equations (\ref{equ:br}) and (\ref{equ:bt}) guarantee divergence-free for the magnetic field in 2D.

In principle, the values for the constants $B_0$, $p$, and $r_0$ should be different from the original fitting results in HQ94 in the modified-BSS field. In HQ94, they fit $p$ and $r_0$ without the $r_\odot/r$ term and the maximum field $B_0$ was set to constant. Meanwhile, in SF83, they obtained a different set of values for $B_0$, $p$, and $r_0$ while taking $r_\odot$=10 kpc and absorbing the sin$p$ factor from $B_r$ to $B_\theta$. Here we simply follow HQ94 to set $B_0=1.4\mu$G\footnote{The field strength around the Sun in HQ94. Note that our modified-BSS field in equations (\ref{equ:br}) and (\ref{equ:bt}) gives a weaker field of 1.0 $\mu$G in the solar vicinity.}, $p=-8.2^\circ$, and $r_0=11.9$ kpc. 

With the new expression, the field strength beyond the main disk becomes weaker due to the newly introduced factor $1/r$. On the other hand, the field strength becomes infinitely high at the Galactic Center (GC), which makes the simulation computationally expensive. In the following section, we introduce a hyperbolic function to smoothly reduce the field strength in the central disk while keeping the modified-BSS field expression in the main disk.

We add in pairs of colliding clouds, each pair being separated by the field reversal. They will collide head-on with the sound speed. The clouds are spherical, each with a radius of 25 pc and a density of $n_H$ = 17 cm$^{-3}$. The cloud density, which is a factor of 10 higher than the uniform disk density, is low compared to typical giant molecular clouds. In fact, they should just be viewed as two overdense spheres because all it needs to trigger CMR are the antiparallel magnetic field, the collision, and the protruding morphology. The spheres may contain atomic gas or molecular gas. As long as the three conditions are satisfied, our goal is to see if CMR can happen in the BSS disk. 


\section{Numerical Method}\label{sec:method}

In this paper, we use the Athena++ code \citep[version 21.0,][]{2020ApJS..249....4S} to model resistive-MHD with a constant Ohmic resistivity ($\eta$). The disk gas is compressible, isothermal, and inviscid. We use the second-order piecewise linear method for spatial reconstruction, with the HLLD Riemann solver for MHD evolution. A second-order van Leer (VL2) time integrator is adopted. No self-gravity is included in the simulation as we focus on the triggering of CMR. For the disk, we include the background potential as in Equation (\ref{equ:Phi}) as a source term in the momentum equation which is integrated explicitly. 

The following equations are solved in our simulations \citep[also see][]{2020ApJS..249....4S}
\begin{equation}\label{equ:mass}
\frac{\partial\rho}{\partial t} + \nabla\cdot(\rho\mathbf{v}) = 0,
\end{equation}
\vspace{-20pt}
\begin{equation}\label{equ:momentum}
\begin{split}
&\frac{\partial\rho\mathbf{v}}{\partial t} + \nabla\cdot(\rho\mathbf{v}\mathbf{v}) = \\
& - \nabla P + \frac{1}{4\pi}(\nabla\times\mathbf{B})\times\mathbf{B} - \rho\nabla\Phi,
\end{split}
\end{equation}
\begin{equation}\label{equ:field}
\frac{\partial\mathbf{B}}{\partial t} = \nabla\times(\mathbf{v}\times\mathbf{B}) + \eta\nabla^2\mathbf{B},
\end{equation}
where $\rho$ is the gas density, $\mathbf{v}$ is the velocity, $P$ is the thermal pressure, $\mathbf{B}$ is the magnetic field, $\Phi$ is the disk potential from Equation (\ref{equ:Phi}), and $\eta$ is the Ohmic resistivity.

The code unit for mass density $\rho$ is set to $3.8\times10^{-23}$ g cm$^{-3}$ (0.50 M$_\odot$ pc$^{-3}$ or $n_H$=17 cm$^{-3}$, assuming a mean mass per H of 1.4). The code unit for time $t$ is set to 20 Myr. The code unit for length scale is 10 pc. The code unit for velocity is 0.51 km s$^{-1}$. With these parameters, the gravitational constant is $G=1$ in code unit. The magnetic field strength unit is 0.31 $\mu$G. The code unit for $\eta$ is 1.5$\times$10$^{24}$ cm$^2$ s$^{-1}$, which is a factor of 10 higher than K21. So if we adopt the same physical value for $\eta$ as K21 (1.5$\times10^{20}$ cm$^2$ s$^{-1}$), the code value for $\eta$ is now 0.0001, which is close to the numerical resistivity ($\la$ 0.0001 based on a test in K21).

\begin{figure}
\centering
\epsscale{1.15}
\plotone{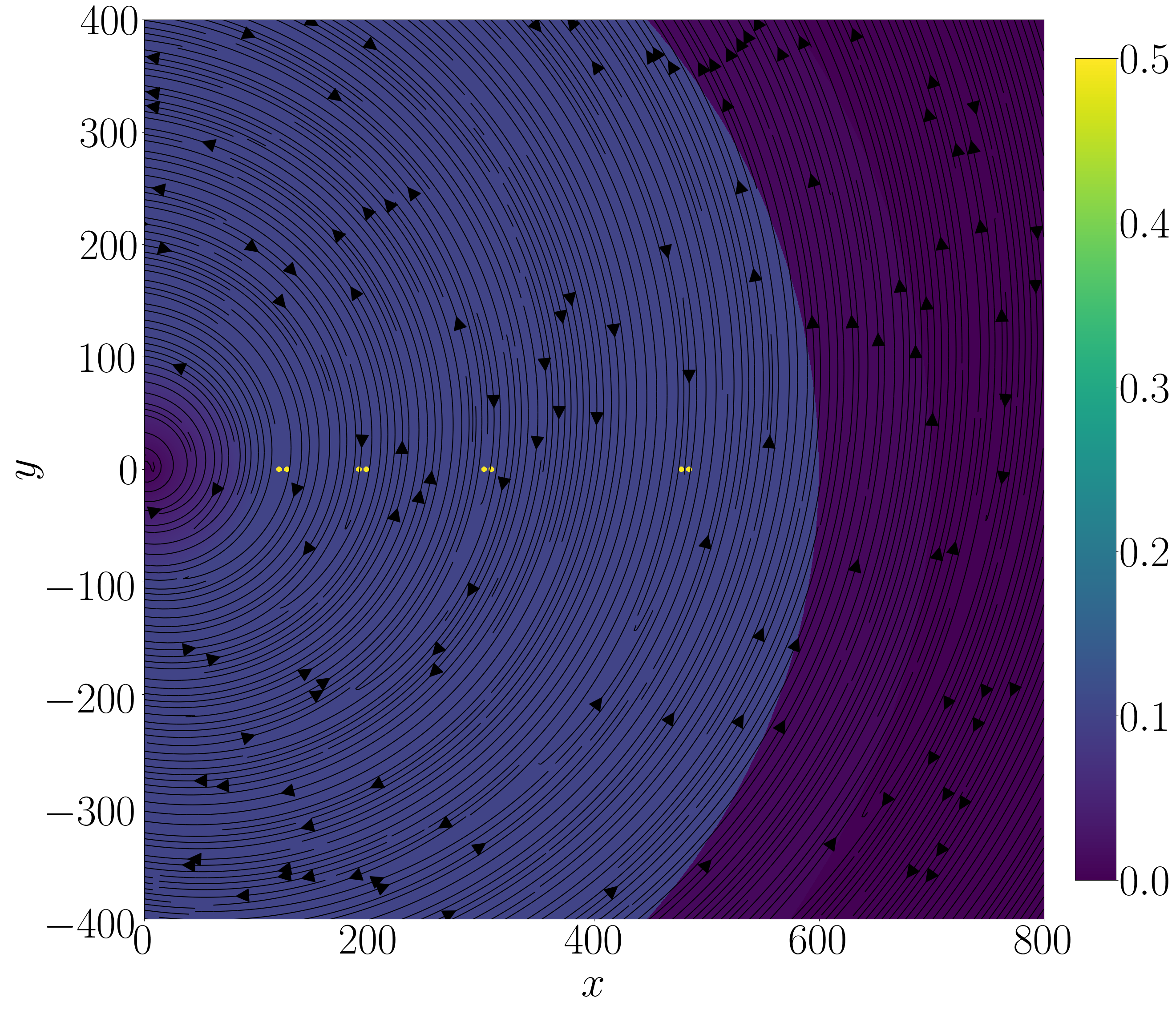}
\caption{
Initial cloud pairs at field reversal interfaces. The color shows the density. The stream lines show the magnetic field. Note the cloud pairs along the positive x-axis. Their density is 1.0.
\label{fig:iccloud}}
\end{figure}

A uniform Cartesian grid is adopted initially with 4096 cells in each dimension, with periodic boundary conditions. To resolve smaller physical scales, we limit the main disk to 6 kpc. The resolution for each dimension is 0.39 (3.9 pc) initially. As shown in K21, high-resolution simulations help CMR in two ways. First, CMR develops dense plasmoids (see discussions later) earlier in higher resolution simulations. Second, gas density in the plasmoids is higher with higher resolution. The fiducial model in K21 (\mbox{MRCOL}) had a grid cell size of 0.0078 pc (a factor of 500 smaller than our initial cell size). We carry out a 2D test of CMR similar to the K21 \mbox{MRCOL2D} model (see their Figure 24) but with a cell size of 0.125 pc, i.e., a factor of 16 worse than K21 but a factor of 32 better than our initial spatial resolution. We do see a density increase at the collision point, implying that CMR happens. But neither the filament nor the plasmoids are resolved and the density increase is only moderate. Then, we increase the resolution by factors of 2 to find out the critical resolution with which the CMR filament becomes prominent. Roughly speaking, the CMR process is significant (the formation of dense gas and helical fields) when the resolution reaches 0.06 pc, i.e., a factor of 8 worse than the fiducial resolution in K21. To reach this cell size, we need to refine our initial uniform grid by a factor of 64 ($2^6$). Therefore, we activate the adaptive mesh refinement (AMR) module in Athena++ with a 6-level refinement. 

Such an AMR simulation is computationally challenging. To save the cost, we set the AMR criteria such that only the region of the cloud-cloud collision is refined. To fulfill this, we require that (1) the curl of magnetic fields is greater than a critical value; (2) the velocity component orthogonal to the field reversal interface is greater than a critical value; (3) the density is greater than a critical factor of the disk density. Criterion (1) selects the field reversal interface where the field curl maximizes. Criterion (2) makes sure there is colliding gas orthogonal to the field reversal. Criterion (3) makes sure there is dense gas production. 

\begin{figure}
\centering
\epsscale{1.15}
\plotone{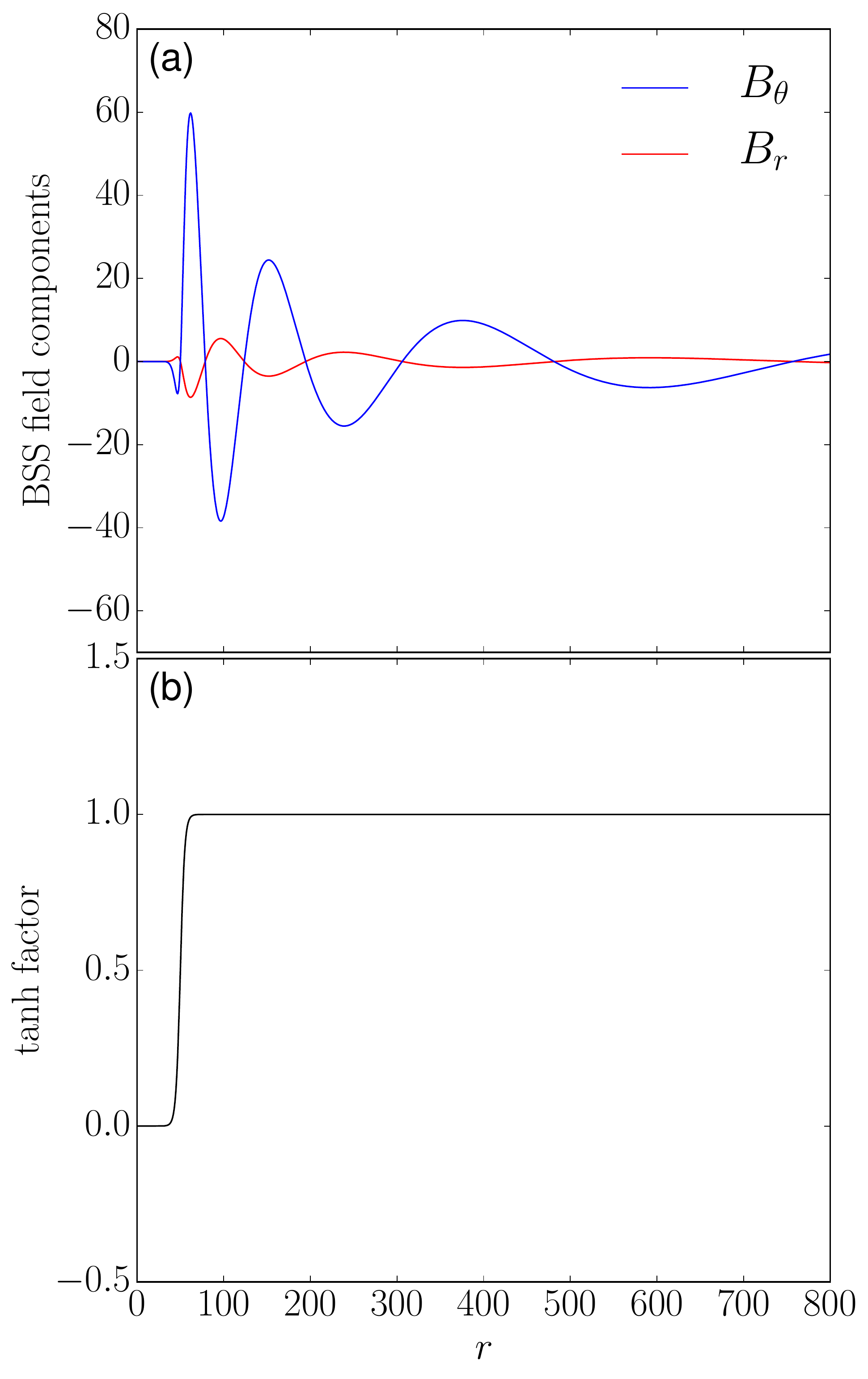}
\caption{
{\bf (a):} The modified-BSS field components as a function of $r$. The field expressions are in equations (\ref{equ:br}) and (\ref{equ:bt}). Here we show the field for $\theta=90^\circ$. The field drops to zero for $r<50$ due to the $tanhfac(r)$ function. 
{\bf (b):} The $tanhfac(r)$ function in equation (\ref{equ:tanhfac}). 
\label{fig:bcurve}}
\end{figure}

Even with these AMR conditions, the computation cost is still high if there are too many refinements. This means that there cannot be too many cloud pairs. For simplicity, we only put cloud pairs at field reversals on the positive x-axis. As shown in Figure \ref{fig:iccloud}, there are four such pairs. Each pair has two spherical clouds. Each cloud has a radius of 25 pc. Each pair of clouds are separated by the field reversal interface. Each cloud has a colliding velocity of the isothermal sound speed (on top of the disk rotation). Hereafter, we name the four collision sites as CMR1, CMR2, CMR3, and CMR4 according to their radii from the disk center. CMR1 is at $R_1\approx$1.2 kpc where each cloud has a mean field of $B_1\approx$1.7 $\mu$G. CMR2 is at $R_2\approx$1.9 kpc where each cloud has a mean field of $B_2\approx$0.75 $\mu$G. CMR3 is at $R_3\approx$3.0 kpc where each cloud has a mean field of $B_3\approx$0.28 $\mu$G. CMR4 is at $R_4\approx$4.8 kpc where each cloud has a mean field of $B_4\approx$0.11 $\mu$G. Because the clouds are near the field reversal interfaces, their field strengths are relatively small. In particular, the fields are a factor of 10-100 weaker than that adopted in K21 fiducial model (10$\mu$G).

With the modified-BSS field in equations (\ref{equ:br}) and (\ref{equ:bt}), the field goes to infinity at $r\rightarrow0$. The strong field in the disk center makes the simulation computationally expensive. In our Milky Way, the field in the GC is complicated. Evidence has shown that the field direction become perpendicular to the disk in the Galactic center, probably transitioning to a part of another large-scale field \citep{2017ARA&A..55..111H}. For instance, the BSS model in SF83 was cut off at an inner radius of 4 kpc.

Because we focus on the main disk, we apply a modified {\it tanh} function that smoothly transitions the field to zero for $r<r_i$ where $r_i$ is a free parameter. For $r>r_i$, the field is unchanged. The modified tanh function is
\begin{equation}\label{equ:tanhfac}
tanhfac(r) = \frac{tanh(\frac{r-r_i}{\delta r})+1}{2}
\end{equation}
where the tanh function is
\begin{equation}\label{equ:tanh}
tanh(x) = \frac{e^x-e^{-x}}{e^x+e^{-x}}
\end{equation}
and $\delta r$ determines how fast $tanhfac(r)$ drops from 1 to 0 from $r>r_i$ to $r<r_i$. We set $r_i=50$ and $\delta r=5$ so that the field drops to zero within the radius of 0.5 kpc. Figure \ref{fig:bcurve}(a) shows the final modified-BSS field. Figure \ref{fig:bcurve}(b) shows the $tanhfac(r)$ function. Note the innermost cloud pair is at 120 (1.2 kpc).

\section{Results and Analysis}\label{sec:results}

\subsection{Fiducial Model}\label{subsec:fiducial}

\begin{figure*}[htb!]
\centering
\epsscale{1.15}
\plotone{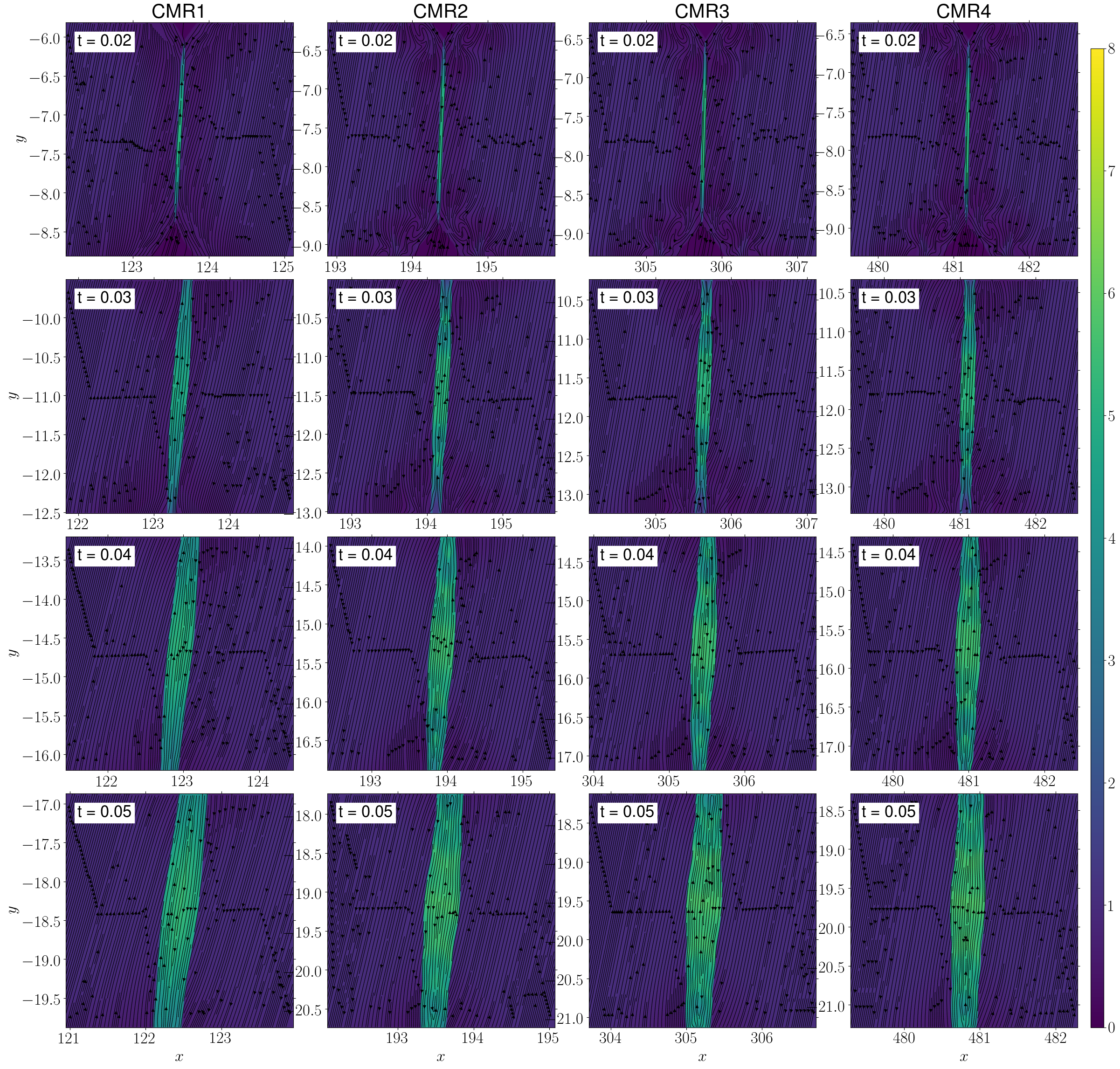}
\caption{
An overview of the fiducial model results. Each panel shows a 3$\times$3 (30 pc$\times$30 pc) region that roughly centers at the collision point. With AMR, the cell scale is 0.006 (0.06 pc). Each column shows one of the four collision sites (CMR1-CMR4). Each row shows a specific time step annotated at the top-left. The color plot shows the gas density while the stream lines show the magnetic field. 
\label{fig:mw0mp1}}
\end{figure*}

Hereafter, we name the fiducial model as MW0. The third column of Table \ref{tab:k21} lists the key parameters of the model. Figure \ref{fig:mw0mp1} shows the collisions between the four pairs of clouds from t=0.02 (0.4 Myr) to t=0.05 (1 Myr). Each panel shows a pair of colliding clouds at a time step. The coordinates are changing due to the disk rotation. The collisions happen at t=0.02 and create compression layers between cloud pairs.

\begin{deluxetable}{cccc}
\tablecaption{Model Parameters \label{tab:k21}}
\tablehead{
\colhead{Model} & 
\colhead{K21} &
\colhead{MW0} &
\colhead{MW1} \\
\colhead{Location} & 
\colhead{-} &
\colhead{CMR1} &
\colhead{CMR1}
}
\startdata
$\delta$ (pc) & 0.0078 & 0.061 & 0.061 \\
$\rho$ (cm$^{-3}$) & 840 & 17 & 17 \\
$T$ (K) & 15 & 8000 & 8000 \\
$B$ ($\mu$G) & 10 & 1.7 & 5.1 \\
$R$ (pc) & 0.9 & 25 & 25 \\
$v_{\rm col}$ (km s$^{-1}$) & 1.0 & 9.0 & 9.0 \\
\hline
$c_s$ (km s$^{-1}$) & 0.29 & 9.0 & 9.0 \\
$v_A$ (km s$^{-1}$) & 0.64 & 0.76 & 2.3 \\
\hline
$\beta$ (2$\rho c_s^2/B^2$) & 0.033 & 23 & 2.5 \\
$R_m$ ($\delta v_{\rm col}/\eta$) & 15 & 1100 & 1100 \\
$S_L$ ($R v_A/\eta$) & $1.1\times10^3$ & $3.7\times10^4$ & $1.1\times10^5$
\enddata
\tablecomments{$\delta$ is the cell size. $\rho$ is the cloud density before collision (in equivalent $n_H$). $T$ is the gas temperature. $B$ is the initial field strength. $R$ is the cloud radius. $v_{\rm col}$ is the collision velocity. 
$c_s$ is the isothermal sound speed. $v_A$ is the Alfv\'en speed $B/\sqrt{4\pi\rho}$. The last three lines list the dimensionless numbers, including the plasma $\beta$, the magnetic Reynolds number $R_m$ (at the scale of grid cell), and the Lundquist number $S_L$. For the Lundquist number, we adopt the cloud radius as the current sheet half-length. MW0 results are in \S\ref{subsec:fiducial} and \S\ref{subsec:force}. MW1 results are in \S\ref{subsec:field}. Both models are further discussed in \S\ref{subsec:k21}.}
\end{deluxetable}

As introduced in \S\ref{sec:intro}, the key feature of CMR is a dense filament at the center of the collision midplane. The filament is wrapped by helical fields. Since our simulation is 2D, we expect to see a density peak at the center of the collision midplane with enclosing field loops. At t$\ga$0.03, we do see field loops in the compression layer. However, the loops remain elongated, failing to pull gas in the layer and create a centrally peaked structure. For t$\ga$0.04 in Figure \ref{fig:mw0mp1}, the compression layer does have a higher density at the center. But the density contrast is not as high as those in K21. Nor do we see magnetic islands (plasmoids).

\begin{figure*}[htb!]
\centering
\epsscale{1.15}
\plotone{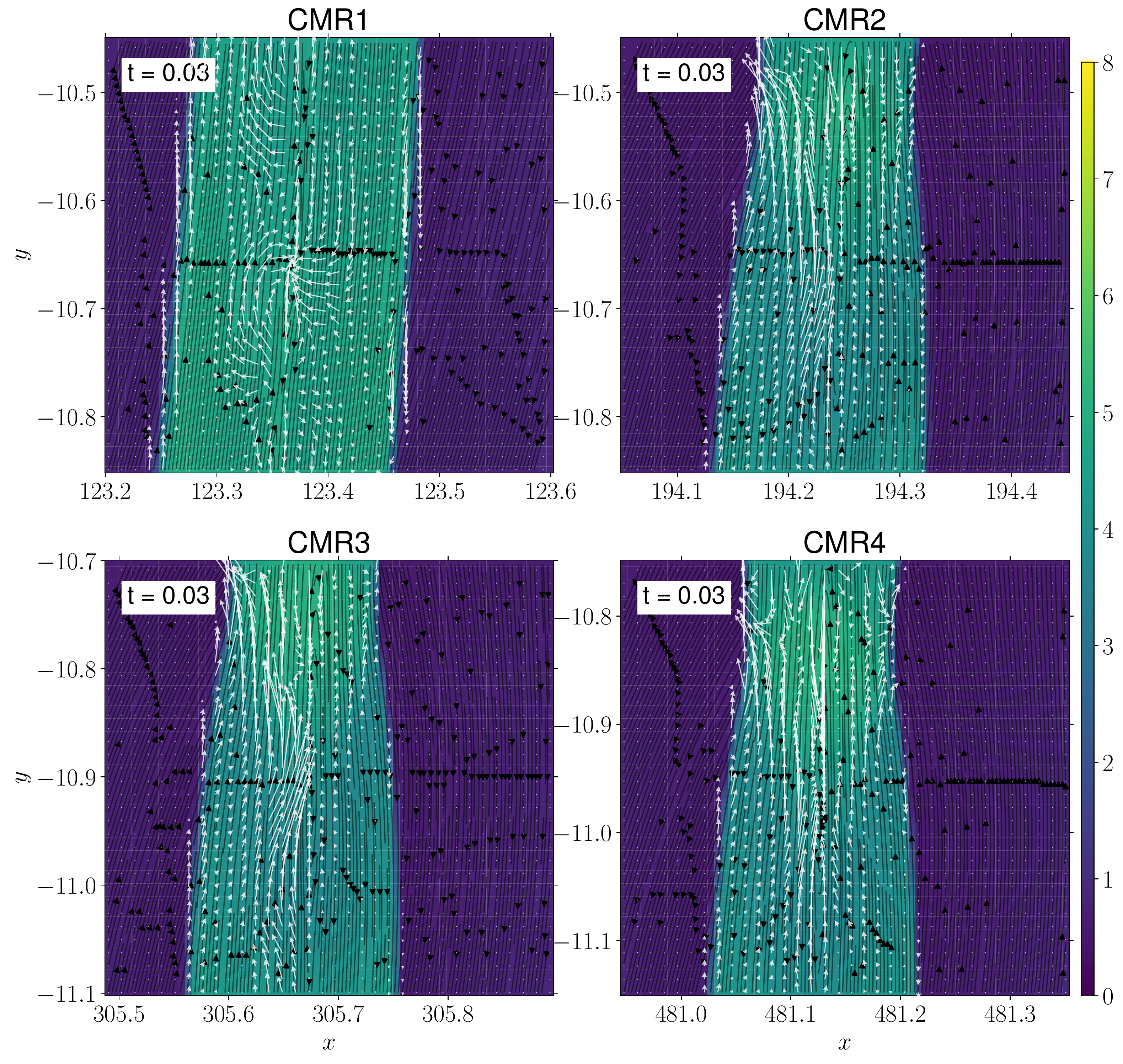}
\caption{
A zoom-in view of magnetic fields at the four collision sites (CMR1-CMR4) at t=0.03 (0.6 Myr). The field-of-view is a 0.4$\times$0.4 (4 pc$\times$4 pc) region centering at field loops. The color plot shows the density while the stream lines show the field. The white arrows show the magnetic tension force. The arrow length is proportional to the force strength within each panel. Between panels, the arrow lengths are not on the same scale.
\label{fig:mw0zmp3}}
\end{figure*}

To see the reconnected field loops and confirm CMR, we zoom into the central 4 pc region of the dense midplane. This field of view has the same size as the computation domain in K21. Figure \ref{fig:mw0zmp3} shows the zoom-in view at t=0.6 Myr. Here, the field streamline separation in both directions is 0.03 pc, i.e., about half of the cell size. For comparison, the streamline separation in Figure \ref{fig:mw0mp1} is 0.25 pc, which is a factor of 4 larger than the cell size.

In Figure \ref{fig:mw0zmp3}, we see field loops forming in all four collision sites. The magnetic tension forces at the top and bottom sides of the loops point toward the loop center, trying to pull gas to the center. The loops and the tension forces confirm the triggering of CMR. This result shows that CMR indeed happens. While we only model four pairs of colliding clouds, there are likely many more colliding clouds along field-reversal interfaces (see Figure \ref{fig:ic}). Therefore, we should expect more CMR events. Since our modified-BSS field is based on studies of our Milky Way, we can address the question from \S\ref{sec:intro} by speculating that CMR should be more common along the Milky Way field-reversal interfaces, which coincide with the spiral arms (HQ94). Note that the BSS field is the large-scale component of the overall magnetic field. There is also small-scale stochastic field which may promote or prevent CMR, partially depending on the net field orientation (K21).

\subsection{Force Comparison}\label{subsec:force}

\begin{figure*}[htb!]
\centering
\epsscale{1.15}
\plotone{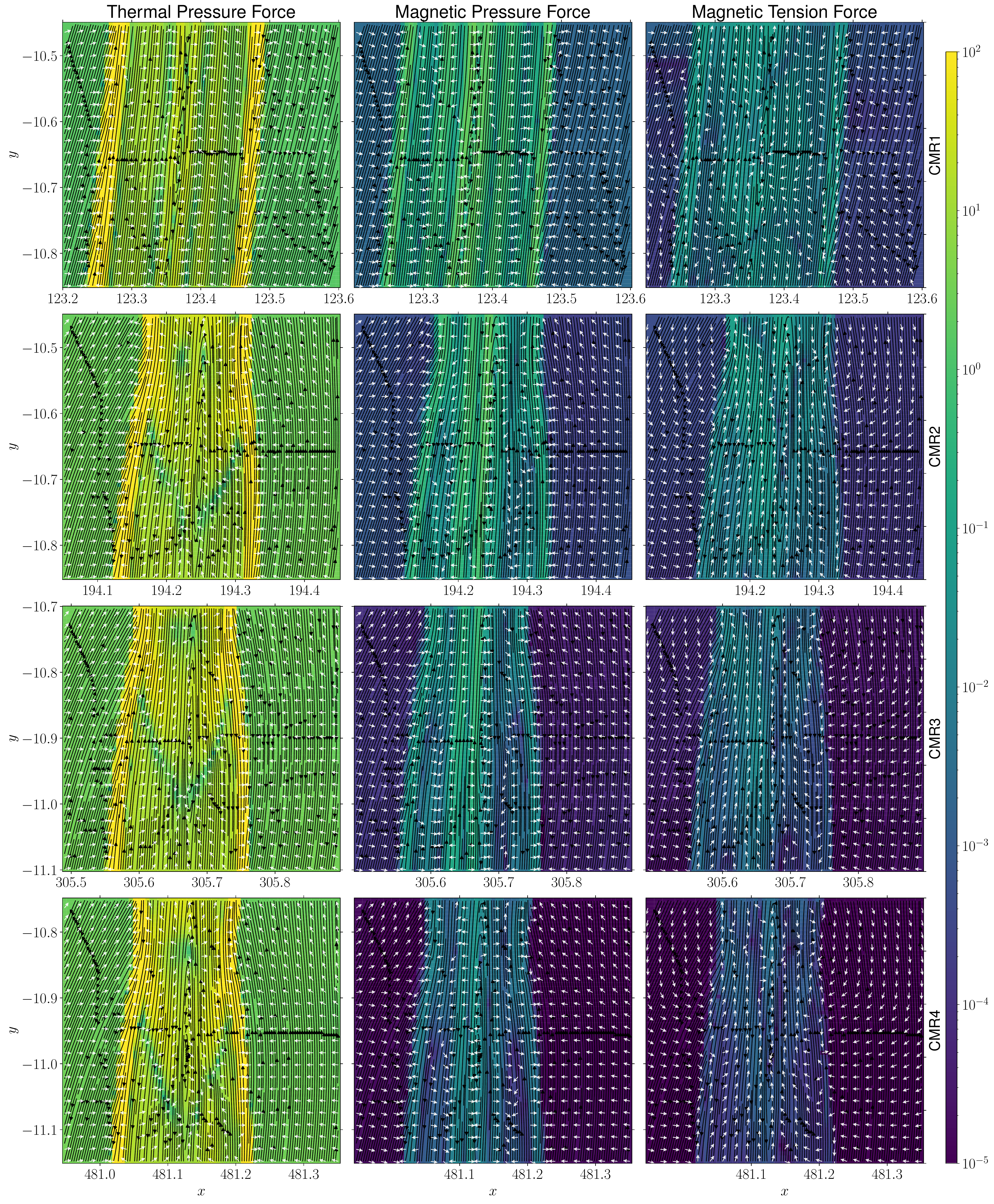}
\caption{
Forces exerted on gas at each collision site. From left to right, each column shows a force term in Equation \ref{equ:force}. From top to bottom, each row shows one of the four CMR sites. The color background shows the force magnitude. All panels share the same scale and limits shown in the right color bar. The white arrows show the force vector directions. The black stream lines show the magnetic fields.
\label{fig:mw0force3}}
\end{figure*}

As shown in Figures \ref{fig:mw0mp1} and \ref{fig:mw0zmp3}, although CMR is triggered at the collision sites, the field loops fail to produce very high density gas. On the contrary, the CMR filament in K21 had a factor of $>$100 density increase. They compared different forces in MHD and showed that the magnetic tension force dominated the other forces. In particular, self-gravity was negligible during the initial filament formation phase. Below, we evaluate contributions from different forces in our model.

The MHD momentum equation (\ref{equ:momentum}) can be rewritten as (ignoring the disk potential)
\begin{equation}\label{equ:force}
\begin{split}
&\frac{\partial}{\partial t}(\rho\mathbf{v}) + \nabla\cdot(\rho\mathbf{v}\mathbf{v}) \\
& = - \nabla P -\nabla\left(\frac{B^2}{8\pi}\right) + {(\mathbf{B} \cdot \nabla)\frac{\mathbf{B}}{4\pi}}.
\end{split}
\end{equation}
It includes three force terms, i.e., the thermal pressure term $-\nabla P$, the magnetic pressure term $-\nabla(B^2/8\pi)$, and the magnetic tension term $(\mathbf{B} \cdot \nabla)\mathbf{B}/4\pi$. Here, $\rho$ is the gas density, $\mathbf{v}$ is the velocity, $P$ is the thermal pressure, and $\mathbf{B}$ is the magnetic field.

As shown in K21, magnetic tension was the dominating force that created the central dense gas. It pulled material from two sides of the elongated field loop to the loop center, overcoming the counteracting thermal pressure force. In our case, however, the magnetic tension is not strong enough. Figure \ref{fig:mw0force3} shows the relative importance of the forces. Each row shows the three force terms for one CMR site. The color scale indicates the force magnitude. All panels have the same color scale for comparison. The arrows indicate the direction of the force. Their lengths are all the same (not proportional to the magnitude). 

Most obviously, the thermal pressure force dominates the other two forces for all four CMR sites. For CMR1, the median value of the ratio between thermal pressure force and magnetic pressure force is $3.8\times10^2$; the median value of the ratio between thermal pressure force and magnetic tension force is $2.6\times10^3$. For the other sites, the ratios are higher due to their weaker fields. As shown in Figure \ref{fig:mw0mp1}, the dense midplane expands at later times due to the overwhelming thermal pressure. While adding self-gravity may help confining the dense gas, the expansion shows that CMR needs to be stronger to make the compressed gas even denser in an environment like the WNM. 

The magnetic pressure force within the compression layer is typically a factor of a few stronger than the tension force. The pressure force is pushing the gas from both left and right. Together, the magnetic forces try to hold the gas in the collision midplane, which shows that CMR still helps maintaining the dense gas.

\subsection{Strong Field Model}\label{subsec:field}

\begin{figure*}[htb!]
\centering
\epsscale{1.15}
\plotone{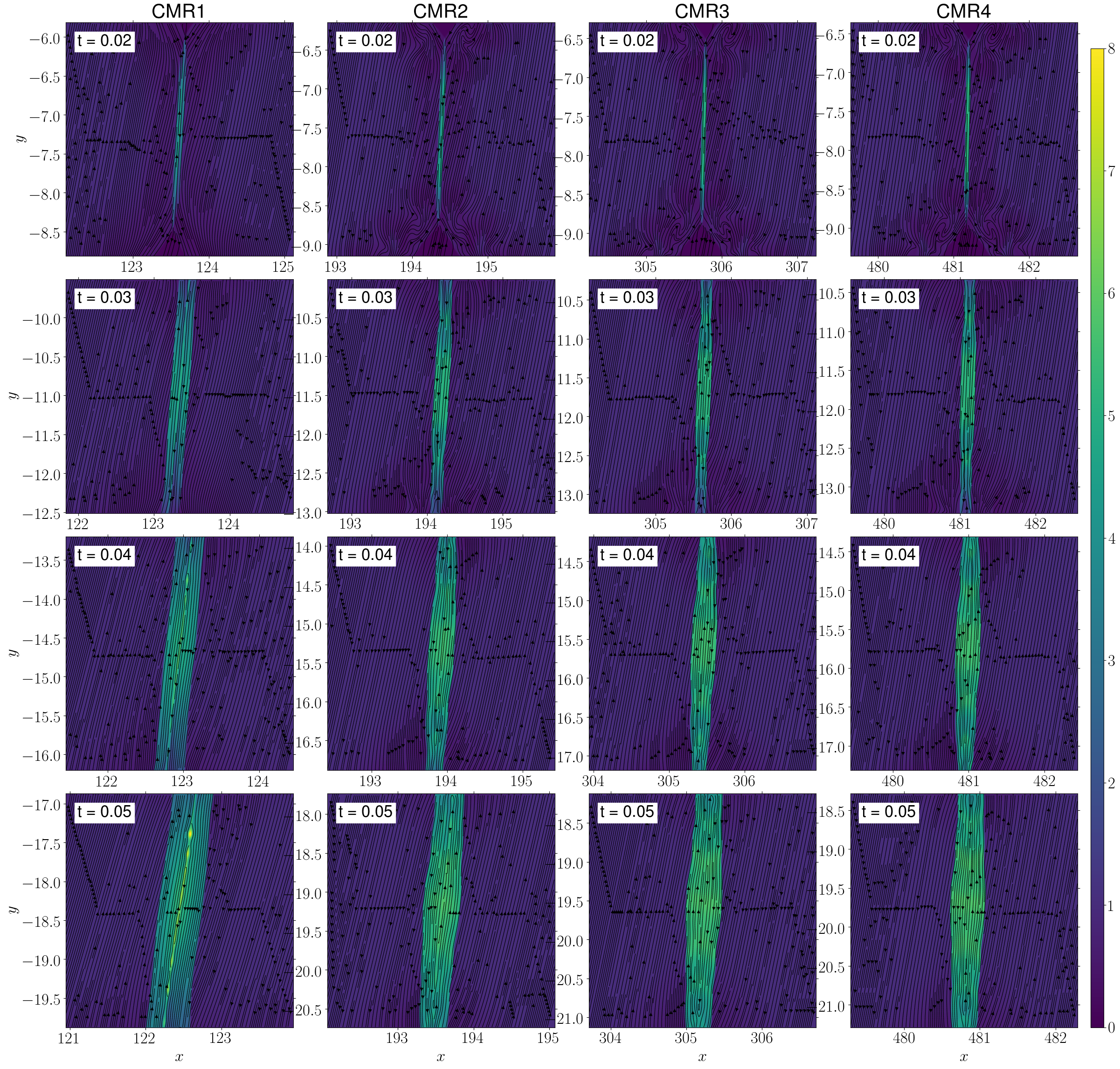}
\caption{
Same as Figure \ref{fig:mw0mp1} but for the strong-field model MW1.
\label{fig:mw1mp1}}
\end{figure*}

Since the magnetic forces in MW0 were dominated by the thermal pressure, a natural idea is to test if stronger fields would enhance the CMR effect. In light of this, we run another simulation with a factor of 3 stronger magnetic field ($B_0=4.2\mu$G). All other parameters remain the same. Hereafter the new model is named as MW1. Figure \ref{fig:mw1mp1} shows the results for MW1. The field strength for CMR1 is $B_1\approx$5.1 $\mu$G, for CMR2 is $B_2\approx$2.3 $\mu$G, for CMR3 is $B_3\approx$0.84 $\mu$G, for CMR4 is $B_4\approx$0.33 $\mu$G. 

Immediately from Figure \ref{fig:mw1mp1} we can see a density enhancement inner layer in the CMR1 collision midplane. From $t=0.02$ to $t=0.03$, all four sites show the compression layer due to the collision. Their densities reach about 5. Starting from $t=0.04$, there is a denser inner layer in the CMR1 compression layer. At $t=0.05$ (1 Myr), the dense layer becomes more prominent. Inside the layer, multiple dense clumps form. Their density reaches an even higher value of $\sim$8 ($n_H=136$ cm$^{-3}$), while the collision compression reaches about 5.7. The formation of the clumps indicates the triggering of the plasmoid instability. The clumps are plasmoids in the current sheet with a density enhancement. 

Meanwhile, the dense clump formation is not seen in the other three CMR sites. Perhaps CMR2 shows a dense inner layer at $t=0.05$ but its density enhancement is not very prominent. Note that the three sites still show field loops indicating magnetic reconnection. The fact that they fail to form dense clumps is probably due to their weaker field strengths. Again, this strength difference is because of our modified-BSS field. Based on the aforementioned field strengths, we anticipate that the critical value for clump formation is $\sim$5 $\mu$G.

As discussed in \S\ref{subsec:fiducial}, CMR can be common along field-reversal interfaces if more collision events happen there. The production of dense gas in MW1 indicates that CMR can be a significant source of dense gas under appropriate conditions. Moreover, the colliding clouds do not have to be molecular since we are just modeling the MHD fluid. As long as they are protruding-shaped structures colliding at the field reversal, we should expect CMR. The structure may be a cloud of ionized or atomic gas. Due to the accumulated column density after CMR, the dense gas may become molecular and form stars. Currently, we only aim to demonstrate the dense gas production by CMR in the disk. We have not included self-gravity. Nor have we run the simulations long enough to see the follow-up gas evolution. Future work will address these aspects step by step.

\section{Discussion}\label{sec:discus}

\subsection{Dissecting the CMR process}\label{subsec:cmr}

The K21 fiducial model was able to form a filament with a $>100$ density increase compared to their colliding clouds. However, in our fiducial model MW0, although magnetic reconnection happens, no dense structure clearly stands out within the compression midplane. Understanding this difference between the two models is crucial to evaluating the importance of CMR in the Galactic disk. 

First, it is necessary to distinguish between magnetic reconnection in a current sheet and fast reconnection due to the plasmoid instability. Generally speaking, there are three phases of magnetic reconnection \citep[see, e.g.,][]{2014ApJ...780L..19P,2016PhPl...23j0702C,2022arXiv220209004J}. In phase-1, antiparallel magnetic fields are brought together to form a current sheet at the field-reversing interface. Here, magnetic reconnection should happen if diffusion becomes important compared to convection (see Equation \ref{equ:field}). Accordingly, the magnetic Reynolds number at the scale of the thickness (or width, $\delta_s$) of the sheet 
$R_{\rm m,s} \equiv \delta_s v_s/\eta$
should be $\la$1. 
Here $v_s$ is the characteristic velocity at the same scale. 
In this phase, the reconnection rate is relatively slow while the thinning of the current sheet continues (the aspect ratio increases).

In phase-2, small plasmoids emerge and enter the linear growth phase $\propto e^{\gamma t}$ as the current sheet thins, where $\gamma$ is the growth rate that is a function of the Lundquist number $S_L\equiv Lv_A/\eta$ ($L$ being the sheet half-length and $v_A\equiv B/\sqrt{4\pi\rho}$ the Alfv\'en velocity). For a steady-state Sweet-Parker current sheet \citep{1957JGR....62..509P,1958IAUS....6..123S}, its aspect ratio ($a/L$ where $a$ is the half-width), growth rate, and mode wave number ($k$) scale with $S_L$ as \citep{tajima1997plasma,2007PhPl...14j0703L}
\begin{equation}\label{equ:sp}
\begin{split}
&a/L \sim S_L^{-1/2},\\
&\gamma\tau_A \sim S_L^{1/4},\\
&kL \sim S_L^{3/8},
\end{split}
\end{equation}
where $\tau_A \equiv L/v_A$ is the Alfv\'en timescale as the outward velocity is typically the Alfv\'en velocity. For large $S_L$, these scaling relations break and the growth rate saturates at the maximum rate \citep{2009PhPl...16k2102B}. The saturation is because the fastest growing mode amplitude exceeds the inner resistive layer with the thinning of the current sheet, when the plasmoid instatibility enters the next phase \citep[a nonlinear growth phase,][]{2016PhPl...23j0702C,2017ApJ...850..142C}. Before that, the plasmoid growth rate not only depends on the Lundquist number but also the mode wave number and the current sheet width. The latter is a function of time because of the sheet thinning \citep{1963PhFl....6..459F,1976FizPl...2..961C}. \citet{2016PhPl...23j0702C} showed that the mode that takes the least time to surpass the linear phase dominates the plasmoid instability.

In phase-3, the plasmoid instability enters a nonlinear growth phase in which fast magnetic reconnection begins. The magnetic reconnection becomes impulsive and bursty, breaking the original current sheet into multiple shorter sheets and secondary plasmoid instabilities occur. If no continuous driving, the system enters a low-energy state by quickly converting magnetic energy to other forms (gas internal energy or energetic particles). Eventually, the reconnection stops. Note, we are not discussing the turbulent reconnection \citep{1999ApJ...517..700L}. Nor do we include turbulence in our simulations. It is a separate topic for future explorations.

However, the CMR mechanism includes more than just the reconnection process itself. Simply speaking, it works through a series of reconnection events initiated by an appropriate geometry. As shown in K21 (see their figure 24), the first two reconnection events happen at the two ends of the compression pancake, with each end ejecting a gas parcel away from the pancake. The ongoing collision (orthogonal to the midplane) creates a field X-point between the pancake end and the parcel. Each X-point thus triggers magnetic reconnection, resulting in two outward reconnected fields. The outward field that faces the compression pancake joins its counterpart from the other end and forms a big field loop enclosing the pancake (also see Figure \ref{fig:cmr}). At this point, two additional processes happen simultaneously. First, the field loop pinches the compression pancake, trying to squeeze it to the center (its symmetry axis). Second, the ongoing cloud-cloud collision continues to compress the pancake. The pancake is essentially a big current sheet and the continuous compression is essentially the sheet thinning process. 

Now, two scenarios for each of the two processes shall follow. First, the field loop can either overcome the thermal pressure in the compression pancake or fail to do so. The outcome is either a filament along the symmetry axis or the largely unchanged pancake. Note that the pancake still has a wrapping field which should show up in observations as a field reversal. Second, the plasmoid instability may or may not develop timely in the compression pancake. Following the above description of magnetic reconnection, the development depends on the Lundquist number, the mode wave number, and the thinning of the pancake. Regardless of the plasmoid instability, the enclosing field loops should bind everything into a single structure if magnetic tension dominates. For example, the high-resolution runs in K21 (see their figure 25) generated many plasmoids along the midplane. They merged as they moved to the center due the enclosing field loops. 

Given the above discussion, we need to consider three separate concepts when comparing our models with the K21 model. The first is whether the two reconnections at the two ends of the compression pancake are triggered and a field loop enclosing the pancake follows. Hereafter we call this the {\it seeding-phase} of CMR. The second is whether plasmoids develop along the collision midplane. Hereafter we call this the {\it plasmoid-phase} of CMR. The third is whether the field loops are able to squeeze the pancake and form a filament. Hereafter we call this the {\it pinching-phase} of CMR. With these definitions, we are in a better position to compare our models with the K21 model.

\subsection{Comparison with K21}\label{subsec:k21}

\begin{figure*}[htb!]
\centering
\epsscale{1.15}
\plotone{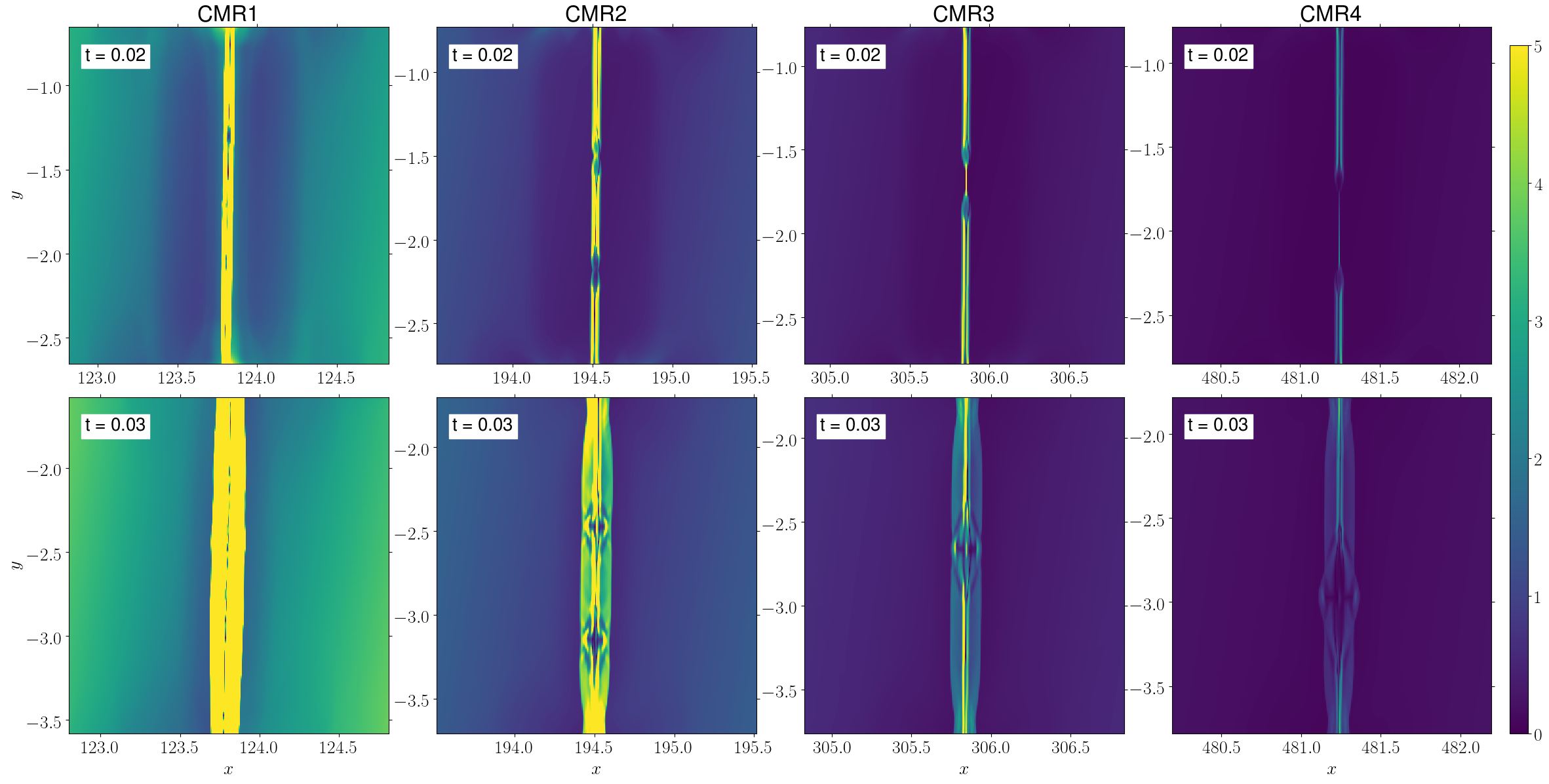}
\caption{
Magnetic field strength $\lvert{\bf B}\rvert$ in color for the four collision sites for MW2 (see \S\ref{subsec:k21}). Each panel shows a $2\times2$ (20 pc $\times$ 20 pc) region. The time step is shown at the top-left corner.
\label{fig:mw2absB}}
\end{figure*}

\begin{figure*}[htb!]
\centering
\epsscale{1.15}
\plotone{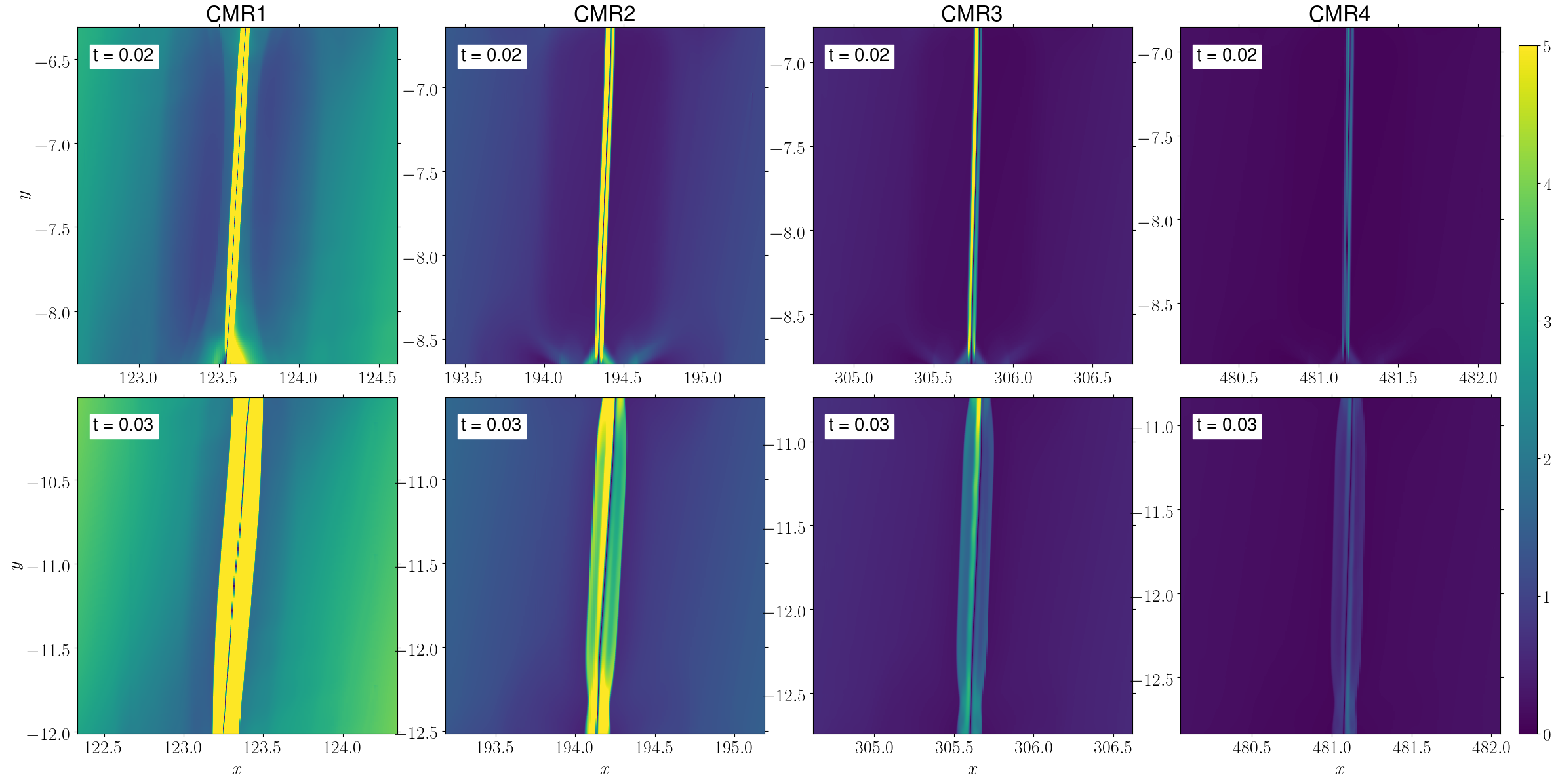}
\caption{
Same as Figure \ref{fig:mw2absB} but for MW0 (see \S\ref{subsec:k21}).
\label{fig:mw0absB}}
\end{figure*}

Because we see field loops in MW0 (\S\ref{subsec:fiducial} and Figure \ref{fig:mw0mp1}) and MW1 (\S\ref{subsec:field} and Figure \ref{fig:mw1mp1}), we believe that the seeding-phase of CMR happens in the two models, which is what we conclude in \S\ref{sec:results}. In the same section, we also notice that the loops fail to squeeze the compression layer into a central dense structure, indicating that the pinching-phase of CMR does not happen. In \S\ref{subsec:force}, we compare the different force terms and find that the critical tension force is not strong enough to dominate the thermal pressure, which is why the pinching-phase fails. This failure is also evidenced by the large plasma $\beta$ in our models compared to that in K21. Table \ref{tab:k21} lists key initial conditions for the K21 fiducial model, our fiducial model MW0 (for CMR1), and our strong field model MW1 (for CMR1). One can see that our models have a plasma $\beta$ that is far greater than that in K21 (the sound speed is superAlfv\'enic).

There is a geometrical artifact that could impact the seeding-phase and the pinching-phase. Although we include AMR, the mesh refinement is performed after the cloud initialization on the coarser base grid.  Consequently, the clouds have a boxy shape. At the beginning of the collision, the contacting geometry is a flat interface instead of a point. Therefore, the first loop, albeit still being created by the seeding-phase, has to confine a larger pancake than that in the K21 model. It is likely taking more time and energy. To make things worse, the follow-up cloud boundaries are like step-functions. So the seeding-phase is like an episodic process. However, the seeding-phase in K21 was more like a continuous process because of the circular geometry of the cloud boundary. So we suspect that the continuous seeding-phase in K21 was able to create a large number of uninterrupted field loops that kept pinching the gradually-growing pancake. Moreover, since we do not include self-gravity, the colliding clouds quickly expand along the field lines and become sparser. They are less capable of overcoming the magnetic pressure and transporting materials to the collision midplane. So it becomes harder to accumulate dense gas.

The plasmoid-phase, however, does not happen in our models, except for CMR1 in MW1. In K21, the plasmoid-phase happened in all of their models (e.g., see their figures 21(a) and 25). Based on the above discussion, we know that our models have experienced the seeding-phase. Therefore, the triggering of magnetic reconnection is present in our models. It should provide an initial perturbation. Below, we provide some qualitative discussion about why we do not see the plasmoid-phase in most of our models.

First, if we focus on K21 and MW0, we see that MW0 has a factor of 8 lower resolution. It is possible that coarser grid cells cannot capture plasmoid modes with large wave numbers. If we consider the steady-state Sweet-Parker context (Equation \ref{equ:sp}), we see that the wave number $k$ is positively related to the Lundquist number $S_L$. MW0 has a factor of 30 higher Lundquist number than K21, which means a plasmoid mode of smaller size scales. So we probably need a factor of 2 higher resolution than K21, meaning another 4 levels of AMR in our simulations. Future work with the desired resolving power that also correctly initializes the spherical cloud morphology should be able to address this issue.

MW0 has a factor of 700 higher plasma $\beta$ than K21 (see Table \ref{tab:k21}). The role of $\beta$ in the onset of plasmoid instability is not fully explored. A recent work by \citet{2012PhPl...19g2902N} investigated the onset of the instability as a function of plasma $\beta$ and Lundquist number $S_L$. They found that the higher the $\beta$, the smaller the critical Lundquist number for the instability. In their case with $\beta=50$, they found a critical $S_L$ of $\sim10^3$. However, in our case, even with $\beta=23$ and $S_L\sim10^4$ in MW0, we do not see the onset of the instability. But we notice that \citet{2012PhPl...19g2902N} assumed an adiabatic equation of state while we assumed an isothermal equation of state. So the thermal energy converted from the magnetic energy is instantly lost. Nonetheless, if we still follow the $\beta-S_L$ trend found by \citet{2012PhPl...19g2902N}, the fact that the K21 model with $\beta=0.033$ and $S_L\sim10^3$ exhibited the plasmoid instability indicates that MW0 should also show the instability. Then, the fact that we do not see the instability implies that $\beta$ is not the cause.

The discussion above also indicates that $S_L$ is probably not the reason MW0 does not show the plasmoid-phase of CMR. However, as we increase $S_L$ by another factor of 3 in MW1, we start to see plasmoids along the collision interface in CMR1 which has the highest $S_L$ among the four collision sites. The plasmoid development indicates an elevated critical $S_L$ for the instability. The elevation is probably not due to the plasma $\beta$, as discussed above.  

A major difference between our model and K21 is that the colliding clouds follow the rotation of the disk. In our fiducial model, the CMR sites are in the flat rotation part of the disk. The rotation velocity is $\sim$200 km s$^{-1}$. Also, gas at different radii feels a differential rotation which tries to tear up the cloud, making it harder for the magnetic tension to pull the dense gas. This effect directly impacts the pinching-phase of CMR. For the plasmoid-phase, it is unclear how the differential rotation promotes or prevents the development of plasmoids. 

We run another test simulation (hereafter MW2) with a factor of 4 slower rotation. All other conditions are the same as MW0. Figure \ref{fig:mw2absB} shows the field strength $\lvert{\bf B}\rvert$ for CMR1-CMR4 sites at $t=0.02$ (0.4 Myr) and $t=0.03$ (0.6 Myr). From the figure, we clearly see plasmoids in all CMR sites. CMR1 has more plasmoids than other sites. But they are smaller than those in other sites. The CMR4 plasmoid is the largest at $t=0.03$. For comparison, Figure \ref{fig:mw0absB} shows the same quantity for MW0. No plasmoids develop in the compression layer. This comparison shows that reducing the disk rotation indeed helps the onset of the plasmoid instability. The pinching-phase of CMR is still absent because of the high plasma $\beta$.

If we re-examine the setup in K21, their initial $\textbf{v}\times\textbf{B}$ were actually curl-free. So the convective term in the induction Equation (\ref{equ:field}) vanished in K21. So in their case, the field diffusion dominated. However, in our case, the convective term dominates the induction equation, which is also manifested by the large magnetic Reynolds number $R_m$ in our models. In Table \ref{tab:k21}, we compute $R_m$ using the collision velocity. $R_m$ becomes a factor of 20 larger if we use the disk rotation velocity. Therefore, it is possible that the disk rotation limits the initial growth of the reconnection (phase-1, see \S\ref{subsec:cmr}). It is also possible that the plasmoid instability is suppressed if the convection rate exceeds the plasmoid growth rate. More explorations are needed in the future.

In fact, the nonequilibrium model in \citet{2016PhPl...23j0702C} indicated that magnetic reconnection depends on the actual thinning process of the current sheet \citep[also see][]{1963PhFl....6..459F,1976FizPl...2..961C}. For our models, the thinning process is largely determined by the cloud-cloud collision and the magnetic field configuration. One of the key features of CMR is that it is a path of current sheet creation in the ISM. Therefore, the inflow velocity to a current sheet originates from the colliding velocity in CMR. On the other hand, the modified-BSS field in our model has an increasing width between reversed fields as a function of increasing radii (see Figure \ref{fig:ic}). So it might be harder for CMR to produce dense gas in the outer disk. An example in \citet{2016PhPl...23j0702C} showed that the fastest mode property depends on the sheet thinning time scale, the Lundquist number, and is a logarithmic function of the timescale, the Lundquist number, the initial sheet width, and the initial perturbation amplitude (see their equation 19). So it is a very complex process. On top of that, one should also consider the differential rotation in a disk environment. However, if the colliding clouds are self-gravitated entities and immune to the differential rotation, CMR may be more prominent.

\section{Summary and Conclusion}\label{sec:conclusion}

In this paper, we have investigated the CMR mechanism in a flat-rotating disk with a modified-BSS magnetic field. Through the investigation, we aim to understand if and how dense gas is produced via the newly discovered mechanism. The outcome is helpful for an initial evaluation of the relative importance of the mechanism. Below we list our itemized findings.

\begin{enumerate}[leftmargin=*]

\item The CMR mechanism is divided into three phases, including the seeding-phase, the pinching-phase, and the plasmoid-phase. The seeding-phase refers to the triggering of the first two reconnection events at the two ends of the compression layer. The pinching-phase refers to the process in which the enclosing field loop pulls the layer to the center due to magnetic tension. The plasmoid-phase refers to the development of plasmoids within the compression layer.

\item In our fiducial model that is representative of WNM, the cloud-cloud collision successfully activates the seeding-phase of CMR, forming field loops within the compression layer. Here, the cloud refers to an overdense region in the WNM disk, not necessarily a molecular cloud. However, the pinching-phase of CMR is not active in that the field loops fail to compress the long compression layer into a central dense structure, which is different from what was originally seen in \citet{2021ApJ...906...80K}. The failure of the pinching-phase is due to the dominance of thermal pressure over magnetic tension. The boxy cloud geometry due to the limited resolution is another possible reason. Nevertheless, the field loops from the seeding-phase still help maintain the dense gas in the compression layer created by the collision.

\item We have run a new model with a stronger magnetic field. Out of the four collision sites, the innermost site with the strongest field develops the plasmoid instability within the central compression layer. Multiple denser clumps emerge along the inner layer of dense gas. The inner layer already has a higher density than the compression layer due to the cloud-cloud collision. Comparing the field strengths among the collision sites, we find a critical value of $\lvert{\bf B}\rvert\ga5\mu$G for the development of the plasmoid-phase of CMR.

\item The above results indicate that CMR can happen in the Galactic disk with a BSS field. It provides a force to help maintain the dense gas from cloud-cloud collision. Under favorable conditions, CMR can create denser gas on top of the collision compression, providing a boost to the dense gas formation in the disk. Moreover, we can imagine more dense gas formation along the field-reversal interfaces in the BSS field if more clouds collide there. Since the interfaces are thought to coincide with the spiral arms of our Milky Way, one can speculate that CMR may be a significant source of dense gas in the Galaxy. 

\item The fact that our strong-field simulation develops plasmoids indicates that our fiducial model has a higher critical Lundquist number for the plasmoid instability in comparison to the models in \citet{2021ApJ...906...80K}. Our qualitative analysis points to our relatively lower resolution and our inclusion of the disk rotation as the causes for the non-development of plasmoids. A test simulation with a slower rotation velocity successfully develops a number of plasmoids for all collision sites, showing another facet of CMR in a rotating environment. We expect that collision between initially self-gravitated clouds may alleviate the rotation effect.
\end{enumerate}

In this paper, we only discuss the role of CMR in a specific configuration of the large-scale magnetic field. In our Milky Way, the stochastic field component is equally important and may induce more CMR events that form dense clouds. If the field is not antiparallel, dense gas can still form \citep{2021ApJ...906...80K} but probably in a more irregular morphology. With cooling and self-gravity, the dense structure can potentially become molecular clouds. More CMR modeling in the future will clarify.

\acknowledgments 
We thank the anonymous referee for an informative report.
SK acknowledges fruitful discussion with Jinlin Han, Mark Reid, Chuanfei Dong, and Luca Comisso.
An allocation of computer time from the UA Research Computing High Performance Computing (HPC) at the University of Arizona is gratefully acknowledged. We thank the Yale Center for Research Computing for guidance and use of the research computing infrastructure, specifically Grace. 

\software{Numpy \citep{numpy}, Matplotlib \citep{matplotlib}, SAOImageDS9 \citep{2003ASPC..295..489J}}



\appendix
\restartappendixnumbering 

\section{Divergence-free BSS Field Expressions}\label{app:bfield}

In HQ94, the expressions for the BSS field is 
\begin{equation}\label{equ:hqbr}
B_r = B_0 \cos \left(\theta-\beta_p\ln\frac{r}{r_0}\right) \sin p,
\end{equation}
\begin{equation}\label{equ:hqbt}
B_\theta = B_0 \cos \left(\theta-\beta_p\ln\frac{r}{r_0}\right) \cos p,
\end{equation}
where the meanings of the quantities are the same as those in equations (\ref{equ:br}) and (\ref{equ:bt}) (note $\beta_p=\cos p/\sin p$). The difference is that HQ94 treated $B_0$ as a constant in their fitting. This choice is not suitable for our simulation setup because the field is not divergence-free. In a 2D cylindrical coordinate system, the divergence of the magnetic field is
\begin{equation}\label{equ:divbhq}
\begin{split}
\nabla\cdot\mathbf{B} &= \frac{1}{r}\frac{\partial}{\partial r}\left(r B_r\right)+\frac{1}{r}\frac{\partial B_\theta}{\partial \theta}\\
&=\frac{1}{r}B_0\cos\left(\theta-\beta_p \ln \frac{r}{r_0}\right) \sin p+\frac{1}{r}\beta_p B_0 \sin \left(\theta-\beta_p \ln \frac{r}{r_0}\right) \sin p-\frac{1}{r}B_0 \sin \left(\theta-\beta_p \ln \frac{r}{r_0}\right) \cos p\\
&=\frac{B_0}{r}\cos\left(\theta-\beta_p \ln \frac{r}{r_0}\right) \sin p.
\end{split}
\end{equation}
For example, with $\theta=0$ and $r=r_0$, $\nabla\cdot\mathbf{B}$ is non-vanishing. Meanwhile, the magnetic field formula from SF83 model A is
\begin{equation}\label{equ:sfbr}
B_r = -B_0\frac{r_\odot}{r}\cos\left(\theta+\beta_p\ln\frac{r}{r_0}\right),
\end{equation}
\begin{equation}\label{equ:sfbt}
B_\theta = B_0\frac{r_\odot}{r}\beta_p\cos\left(\theta+\beta_p\ln\frac{r}{r_0}\right),
\end{equation}
which is divergence-free. However, the $\beta_p$ term in the azimuthal component in equation (\ref{equ:sfbt}) makes the field magnitude very high. Instead of giving the $\sin p$ term to the radial component and the $\cos p$ term to the azimuthal component (as in HQ94), SF83 merged the two terms into the $\beta_p$ term in the azimuthal component. Due to the small pitch angle (SF83 had -5$^\circ$), $\beta_p$ is large (-12 in SF83). So the field magnitude for inner disk is relatively high. SF83 only fit their model for $r>4$ kpc. To solve these issues while largely maintain the BSS model, we use the following modified-BSS model 
\begin{equation}\label{equ:kbr}
B_r = B_0\frac{r_\odot}{r}\cos\left(\theta-\beta_p\ln\frac{r}{r_0}\right)\sin p,
\end{equation}
\begin{equation}\label{equ:kbt}
B_\theta = B_0\frac{r_\odot}{r}\cos\left(\theta-\beta_p\ln\frac{r}{r_0}\right)\cos p,
\end{equation}
where we only include a minor change to the HQ94 expressions. It can be shown that the modified expressions are divergence-free:
\begin{equation}\label{equ:divbk}
\begin{split}
\nabla\cdot\mathbf{B} &= \frac{1}{r}\frac{\partial}{\partial r}\left(r B_r\right)+\frac{1}{r}\frac{\partial B_\theta}{\partial \theta}\\
&=\frac{r_\odot}{r^2}\beta_p B_0 \sin \left(\theta-\beta_p \ln \frac{r}{r_0}\right) \sin p-\frac{r_\odot}{r^2}B_0 \sin \left(\theta-\beta_p \ln \frac{r}{r_0}\right) \cos p\\
&=0.
\end{split}
\end{equation}

Note that the modified-BSS model is solely for the consistency in 2D simulations. It does not invalidate the HQ94 expressions because in 3D there is a z-direction field component that potentially keeps the 3D field divergence-free.

\bibliography{ref}
\bibliographystyle{aasjournal}

\end{document}